\documentclass[preprint,pre]{revtex4-1}
 
\usepackage{graphicx}
\usepackage{amsmath}
 
\usepackage{amssymb}
\usepackage{natbib}
  
\usepackage{color}

\begin{document}

\newcommand{\inl}[1]{$ #1 $}

\title{Activity induced collapse and re-expansion of rigid polymers}
\author{J. Harder$^1$,  C. Valeriani$^{2}$, A. Cacciuto$^1$}
\email{ac2822@columbia.edu}
\affiliation{$^{1}$Department of Chemistry, Columbia University\\ 3000 Broadway, New York, NY 10027\\ }
\affiliation{$^{2}$Departamento de Quimica Fisica, Facultad de Ciencias Quimicas, Universidad Complutense de Madrid, 28040 Madrid, Spain}

\begin{abstract}
We study the elastic properties of a rigid filament  in a bath of self-propelled particles.
We find that while fully flexible filaments swell monotonically upon increasing the strength of the propelling force,  rigid filaments soften for moderate activities, collapse into metastable hairpins  for intermediate strengths,  
and eventually re-expand  when the strength of the activity of the surrounding fluid is large. This collapse and re-expansion of the filament with the bath activity is reminiscent of the behavior observed in polyelectrolytes in the presence of different concentrations of multivalent salt. 
\end{abstract}

%%\keywords{latex-community, revtex4, aps, papers}

\maketitle

\section{Introduction}
When a micron-sized object is immersed in a thermally equilibrated solution, it will undergo Brownian motion due to random collisions with the surrounding fluid. The situation is vastly different when this object is suspended in
a bath of active (self-propelled) particles, especially when its shape is not spherical. Indeed, it is well known that  
self-propelled  particles tend to accumulate around highly curved regions of suspended obstacles, thus generating density local gradients which can lead to effective propulsive forces, and anomalous rotational and translational diffusion. Recent work has explored the trapping of active particles on boundaries and microstructures of different shapes and how
passive obstacles can become activated in the presence of an active bath
~\cite{tailleur,morozov,kaiser2014,wan2008,kaiser2012,kaiser2013,angelani2010,wensink2014,wu2000,elgeti2013,mallory2014,wensink2008,leptos2009}. More recently, we have explicitly characterized how the curvature of passive tracers is  related to the extent of their activation~\cite{cacciuto2014}. Needless to say, having a way of selectively activating passive microstructures  and their aggregates based on their shape is the first step towards the fabrication of microscopic engines, where the random active motion of one component can be converted into the persistent motion of another. A beautiful example of this process was recently shown in~\cite{angelani2009,dileonardo2010}.

In this paper, we go one step further and consider the case of a passive structure with a flexible shape.
Specifically, we study  a semi-flexible polymer (filament) confined in two-dimensions in
the presence of a low concentration of active spherical particles. Our analysis discusses the behavior of the filament for different values of the strength of the bath activity (characterized by the particles' self-propelling force) and of the filament rigidity.
Recent work on active flexible filaments (i.e.~chains formed by active Brownian particles) has  shown very interesting phenomenological behavior, including periodic beating and rotational motion~\cite{Chelakkot}. While preparing this manuscript, Kaiser et al.~\cite{Kaiser} have considered a system similar to ours, but in the fully flexible limit. In their study they show how a filament undergoes  anomalous swelling upon increasing activity of the bath.
Here, we confirm the results  reported in~\cite{Kaiser}, and show  how adding rigidity to the 
filament leads to an even more complex non-monotonic behavior, where the chain
softens for moderate activities, can collapse into a metastable hairpin  for intermediate activities, and 
behaves as a fully flexible chain in the large activity limit. 
Our data also indicate that at intermediate activities a rigid filament behaves as a two state system that  dynamically breathes between a tight hairpin and an ensemble of extended conformations.

\section{Methods}
We perform numerical simulations of a self-avoiding polymer confined in a two-dimensional plane and immersed in a bath of spherical 
self-propelled particles of diameter $\sigma$, whose motion is described via the coupled Langevin equations 
\begin{align}
m \ddot{ \boldsymbol{r}}  & =-\gamma \dot{ \boldsymbol{r}}+F_a{\boldsymbol n}(\theta)-\partial_{\boldsymbol r} V+\sqrt{2\gamma^2D} {\boldsymbol \xi}(t)  \\
 \dot{\theta}& =\sqrt{2 D_r} \,\,\xi_r(t) 
\end{align}
\noindent Here, $\gamma$ is the friction coefficient,  $V$ the pairwise interaction potential between the particles,  $D$ and $D_r$ are the translational and rotational diffusion constants respectively, satisfying the relation $D_r=3D/\sigma^2$. ${\boldsymbol n}(\theta)$ is a unit vector pointing along the propelling axis of the particles, and its orientation, tracked by the angular variable $\theta$, and is controlled by the simple rotational diffusion in Eq 2.  $F_a\geq 0$  is the magnitude of the self-propelling force acting on each particle. The  solvent-induced Gaussian white noise terms for both the translational and rotational motion are characterized by the usual relations $\langle \xi_i(t) \rangle = 0$,   $\langle \xi_i(t) \cdot   \xi_j(t') \rangle = \delta_{ij}\delta(t-t')$ and $\langle  \xi_{r}(t)\rangle = 0$, $\langle  \xi_{r}(t) \cdot   \xi_{r}(t') \rangle =\delta(t-t')$. 
The polymer consists of $N=70$ monomers of diameter $\sigma$, linearly connected via harmonic bonds according to the potential $ U_b = \varepsilon_b\left(r - r_0\right)^2 $ where $r$ is the center-to-center distance between linked monomers, $\varepsilon_b = 10^3k_{\rm B}T/\sigma^2$, and $r_0 = 2^{1/6}\sigma$ is the equilibrium distance.
An additional harmonic potential, $U_a = \kappa\left(\phi - \phi_0\right)^2$, is used to introduce angular rigidity to the polymer, where $\phi$ is the angle between  adjacent bonds, and $\phi_0 = \pi$ is the equilibrium angle.  Values of $\kappa$ considered in this work range from $0$ to $90k_{\rm B}T$. The equation of motion of the filament monomers are given by
Eq 1., with $F_a=0$.

The excluded volume between any two particles in the system (including the monomers in the chain) is enforced with a  purely repulsive  Weeks-Chandler-Andersen  potential
\begin{equation}
 V(r)=4 \epsilon \left[ \left( \frac{\sigma}{r} \right)^{12}- \left( \frac{\sigma}{r} \right)^{6}+\frac{1}{4} \right]  
 \label{eq:LJ_V}
\end{equation} 
\noindent with range extending up to $r=2^{1/6}\sigma$. Here $r$ is the 
center-to-center distance between any two particles, and $\varepsilon=10\,k_{\rm B}T$. 
The simulation box is a square with periodic boundary conditions. All simulations have been 
performed using LAMMPS\cite{Plimpton1995} at an active particle volume fraction $\phi=0.1$ (corresponding to 318 active particles), which is sufficiently small to prevent  spontaneous phase separation of the active particles~\cite{marchetti2,catesx}. 
Throughout this work we  use the default dimensionless Lennard Jones units as defined in LAMMPS, for which the fundamental quantities mass \(m_0\), length \(\sigma_0\), epsilon \(\epsilon_0\), and the Boltzmann constant \(k_{B}\) are set to 1, and all of the specified masses, distances, and energies are multiples of these fundamental values. In our simulations we have $T=T_0 = \epsilon_0/k_{B}$, $m=m_0$, $\sigma=\sigma_0$.  Finally $\gamma=10\tau_0^{-1}$ where $\tau_0$ is 
the dimensionless unit time defined as \(\tau_0 = \sqrt{\frac{m_0 \sigma_0^2}{\epsilon_0}}\).

\section{Results \& Discussion}
\begin{figure}[!h]
\centering
\includegraphics[scale = 1]{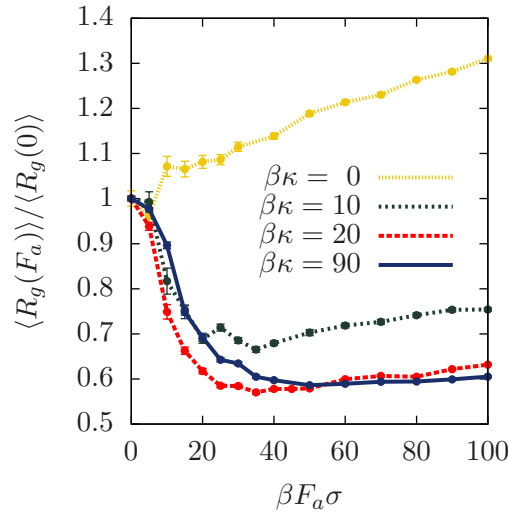}
\caption{Time averaged radius of gyration of a filament, normalized by its value calculated in the absence of activity ($F_a=0$), as a function of the propelling force $F_a$ for different filament rigidities, $\kappa$.}
\label{fig:kick_v_stiff}
\end{figure}

We begin our analysis by measuring the radius of gyration of the filament, $R_g^2 = \frac{N^2}{2} \sum_{i,j}\left(\mathbf{r_i} - \mathbf{r_j}\right)^2$ as a function of the strength of the propelling force of the surrounding particles in the bath, $F_a$, for different values of the chain angular rigidity $\kappa$. The results are shown in Fig. \ref{fig:kick_v_stiff}.

Consistent with the work by Kaiser et al.~\cite{Kaiser}, a fully flexible self-avoiding filament swells when immersed in an active bath. This is due to the forces exerted on the filament by the active particles as they cluster along different regions of the filament and  stretch it by pushing it in different directions.
Interestingly, when the filament is somewhat rigid,
the opposite behavior is observed. The radius of gyration 
decreases for small values of $F_a$, it reaches a minimum for intermediate activities, and it slowly begins to increase for large propelling forces. This non-monotonic behavior  of the radius of gyration with $F_a$ is reminiscent of that of polyelectrolytes in the presence of multivalent salt~\cite{luijten} as a function of salt concentration.

When an active particle strikes a fully flexible filament, the filament can be stretched from a relatively compact configuration  to a more extended one.
However, when the filament is stiff, it is mostly stretched to begin with, and any local asymmetries in the number of active particles that cluster on either side of the filament can provide sufficiently large forces to bend it.
As the filament begins to bend (say in the middle), different regions of its surface will have different exposure to the active particles. Specifically, the outer region (see Fig. 2(a)) will experience a larger number of collisions than the inner region as the former offers a larger cross section to the active particles. Simultaneously, active particles in the inner region tend to accumulate around the areas with highest curvature [14]. The combination of the forces acting on the inner and outer regions of the filament creates a pivot around which the filament can fold 
into a tight hairpin. 
  
\begin{figure}[!h]
\centering
\includegraphics[scale = 1]{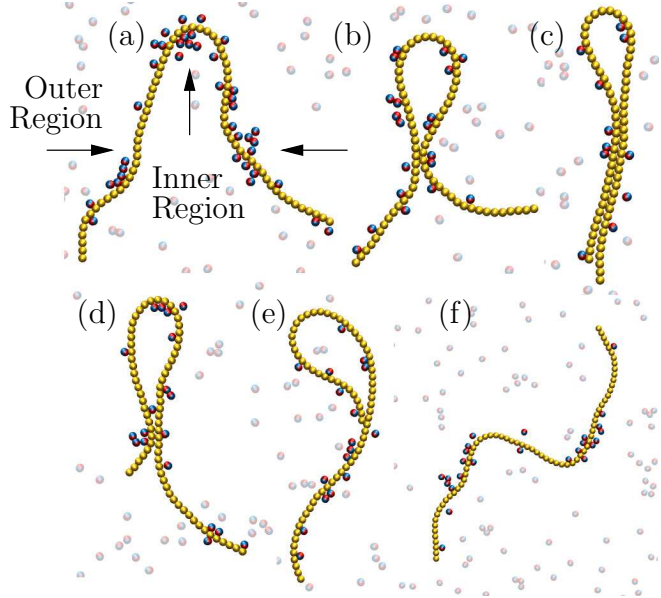}
\caption{Snapshots from our simulations depicting the filament at various stages of folding and unfolding (a)$\rightarrow$(f). The direction of the propelling force $F_a$  follows the axis piercing the poles of the two hemispheres we used to depict the active particles in the blue-to-red direction. Panel (a) shows explicitly our convention for the definition of inner and outer regions of a bent filament. For the sake of clarity, active particles away from the filament have been rendered with a semi-transparent filter.\cite{VMD}}
\label{hairpin}
\end{figure}

Because there are no direct attractive interactions between different parts of the filament or between the filament and the active particles, the hairpin eventually unfolds. This process is initiated by the active particles located in the inner region whose forces cause the location of the hairpin's head to slide along the filament, effectively shortening one arm and lengthening the other until full unfolding is achieved. Fluctuations in the number of active particles applying a force on the outer arms of the filament allow it to open slightly, thus letting active particles either out of the inner region or into it which stabilizes or de-stabilizes the filament, respectively.
For a range of intermediate bath activities, the filament will therefore breath dynamically between a tightly folded and an ensemble of bent but extended configurations.  
Fig.~\ref{hairpin} shows several snapshots from our simulations highlighting the 
the folding and unfolding of the filament.

\begin{figure}[!h]
\centering
\includegraphics[scale = 1]{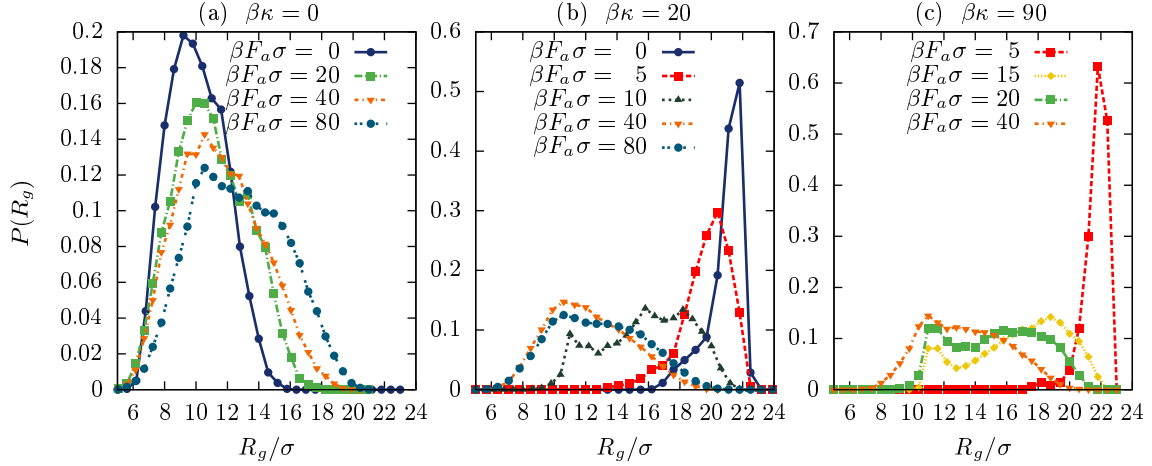}
\caption{Probability distribution of the radius of gyration  $P(R_g)$  for increasing values of the propelling forces $F_a$. 
Panel (a) shows the results for a fully flexible chain, panel (b) those for a polymer with rigidity $\beta\kappa=20$, 
and panel (c) those for the stiffest  polymer considered in this work, $\beta\kappa=90$.}
\label{fig:rg_hist}
\end{figure}

Given this complex dynamics, the average value of the filament radius of gyration
provides only a limited characterization of the system. Crucial
information about the conformation of the filament can be extracted from the probability distribution, $P(R_g)$, of the radius of gyration, for different values of $\kappa$ and $F_a$. 
For a fully flexible chain, Fig.~\ref{fig:rg_hist}(a), we observe that upon increasing the bath activity,
the distribution, $P(R_g)$, begins to broaden towards larger values of $R_g$, indicating that the filament is able to reach configurations that are more extended (swollen) with respect to the typical behavior expected in the absence of active particles,
%{\bf For very large values of $\beta F_a\sigma$, $P(R_g)$ develops a peak around a radius of gyration compatible with a fully stretched configuration, and the distribution presents a fat tail that includes configurations with significantly smaller $R_g$.}
which is consistent with  recent results by Kaiser et al. 
For rigid filaments, Fig. \ref{fig:rg_hist}(b)-(c), we observe the opposite behavior  as a function of the bath activity.
Namely, the sharply peaked distribution for large values of $R_g$ that one would expect in the absence of particle activity and that corresponds to a fully stretched filament, broadens and shifts towards lower values of $R_g$ as $F_a$ is increased in the systems. This is indicative of the  softening of the chain. Upon further increase of $F_a$,   distinct peaks begin to develop (signature of the formation of folded conformations). Finally, for even larger values of $F_a$, we observe a broad distribution covering a wide range of $R_g$.
Strikingly, the distributions in the large activity limit are independent of $\kappa$,
implying that that a rigid filament in a bath at large activity behaves as a fully flexible filament in the same bath.  

Further evidence  for the formation of hairpins and their relative stability with respect to  more extended configurations can be obtained by combining calculations of  $R_g$ with the corresponding information about the asphericity of the filament. Following Rudnick et al.~\cite{Rudnick} we define asphericity as:
\begin{equation}
\langle A \rangle = \frac{\sum_{i>j}\langle\left(\lambda_i-\lambda_j\right)^2\rangle}{\langle\left(\sum_i\lambda_i\right)^2\rangle}
\end{equation}
where $\lambda_i$ is the $i^{\mathrm{th}}$ eigenvalue of the shape tensor.  In two dimensions, this parameter is equal to $0$ for a circular or isotropic object, and $1$ for an object which extends along one dimension.  In our system, small values represent configurations where the filament is relatively symmetric about its center of mass, whereas large values indicate a fully stretched, rigid filament. 

\begin{figure}[!h]
\centering
\includegraphics[scale = 1]{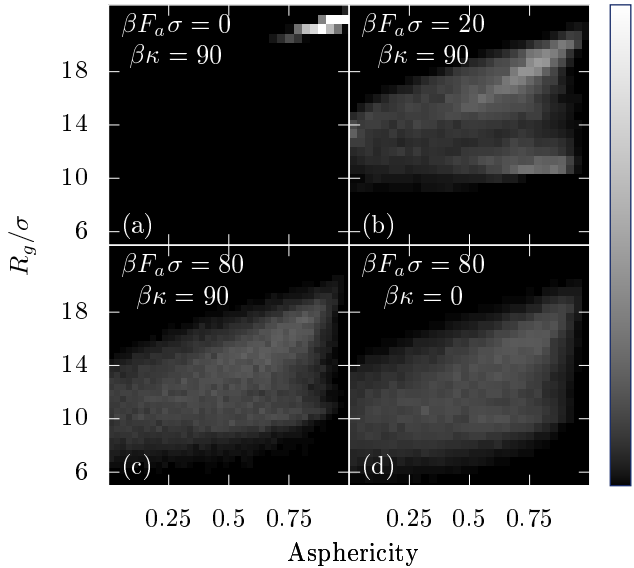}
\caption{Panels (a)-(c). Time averaged probability distribution of finding a rigid filament, $\beta\kappa=90$, with a given radius of gyration $R_g$ and asphericity $A$ for different values of $F_a$.  When $\beta F_a\sigma=0$, panel (a), the  filament is almost exclusively extended ($A\sim 1$, $R_g\sim 22\sigma$).  At $\beta F_a\sigma = 20$, panel (b), two distinct peaks can be seen at large values of the asphericity, where one has roughly half the $R_g$ of a fully extended chain.  These peaks correspond to the folded (hairpin) and unfolded (extended) states.  At larger values of $F_a$, panel (c), the distribution smears out and the filament can more freely explore a wider variety of configurations. Note how the distribution in panel (c) is very similar to that in panel (d) corresponding to a fully flexible filament, $\beta\kappa=0$, in an active fluid having the same propelling force $\beta F_a\sigma = 80$. }
\label{fig:rg_asphere}
\end{figure}

We show the result of this analysis for $\beta\kappa=90$ in Fig. \ref{fig:rg_asphere},
where we plot the  time averaged probability distribution of configurations  with a given $R_g$ and asphericity, $A$. When $\beta F_a\sigma=0$ the plot shows, as expected, a distribution centered
around large values of $R_g$ and $A$ close to 1, corresponding to a stiff filament.
As the bath activity increases, at $\beta F_a\sigma=20$ two distinct high probability regions emerge.
Both have large asphericities, indicating elongated configurations, but with well
separated values of $R_g$.  The folded configuration has a value of $R_g \sim 11\sigma$ and the unfolded corresponds to $R_g \sim 18\sigma$.  This scenario, consistent with our previous analysis,
suggests a two-state system (folded-unfolded) where the two states have comparable probabilities. As $F_a$ is further increased, the distribution smears out as local folding and unfolding of filament segments begin to occurs at every length-scale along the chain and the filament's statistics begin to resemble that of a fully flexible swollen polymer. For comparison we also show in the same figure (bottom-right) the probability distribution for a fully flexible filament immersed in a bath with the same active force.
 
\section{Conclusions}
In this paper we studied the behavior of a rigid filament immersed in a low volume fraction suspension 
of active particles  confined in two dimensions as a function of filament rigidity and bath activity. 
Using shape-sensitive observables,
we fully characterize the conformational properties of the filament.
Our findings indicate that fully flexible filaments swell monotonically with bath activity, but rigid filaments present a more peculiar non-monotonic behavior. For small activities, rigid filaments soften and become more flexible. As the bath activity is further increased, we observe the formation of a two state system characterized by extended and hairpin configurations, with complex local dynamics leading to their formation and de-stabilization.
Finally, in the large bath activity limit the behavior of fully flexible and rigid filaments become nearly identical with a large distribution of accessible filament conformations.
It should be stressed that this behavior is peculiar to two-dimensional or quasi-two dimensional confinement.
In fact, simulations of the same system in three dimensions (not presented) do not show traces of the 
remarkable phenomenology observed in two dimensions. This is because of the  reduced probability of a collision 
between active particles and filaments, but more importantly because of the inability of a one-dimensional filament to effectively trap active particles in three dimensions. We however expect a  behavior similar to that observed for polymers in two-dimensional membranes embedded in a three dimensional bath of active particles. Work in this directions will be published elsewhere.

\section*{Acknowledgements}
\indent AC and CT acknowledge financial support from the National Science Foundation under Grant No. DMR-1408259. CV acknowledges financial support from a Juan de la Cierva Fellowship, from the Marie Curie Integration Grant 322326-COSAAC-FP7-PEOPLE-CIG-2012, and from the National Project FIS2013-43209-P.

%\bibliographystyle{apsrev4-1}
%\bibliography{polymer_active_10_9}


\end{document}